\documentclass[a4paper,12pt]{article}
\usepackage{amsmath}
\usepackage{amssymb}
\usepackage{amsfonts}
\usepackage{graphics}
\usepackage{epsfig}

\begin{document}
\renewcommand{\theequation}{\thesection.\arabic{equation}}
\thispagestyle{empty}
\vspace*{-1.5cm}

\begin{center}
{\Large\bf  Towards a quantum field theory of primitive string fields\\}
\vspace{2cm}
{\large Werner R\"uhl}\\
Department of Physics, Technical University of Kaiserslautern\\P.O.Box 3049,
67653 Kaiserslautern, Germany \\
\vspace{2cm}
\begin{abstract}
We denote generating functions of massless even higher spin fields "primitive string fields" (PSF's). In an introduction we present the necessary definitions and derive propagators and currents of these PDF's on flat space. Their off-shell cubic interaction can be derived after all off-shell cubic interactions of triplets of higher spin fields have become known [2],[3]. Then we discuss four-point functions of any quartet of PSF's. In subsequent sections we exploit the fact that higher spin field theories in $AdS_{d+1}$ are determined by AdS/CFT correspondence from universality classes of critical systems in $d$ dimensional flat spaces. The $O(N)$ invariant sectors of the $O(N)$ vector models for $1\leq N \leq \infty$ play for us the role of "standard models", for varying $N$, they contain e.g. the Ising model for $N=1$ and the spherical model for $N=\infty$. A formula for the masses squared that break gauge symmetry for these $O(N)$ classes is presented for d = 3. For the PSF on $AdS$ space it is shown that it can be derived by lifting the PSF on flat space by a simple kernel which contains the sum over all spins. Finally we use an algorithm to derive all symmetric tensor higher spin fields. They arise from monomials of scalar fields by derivation and selection of conformal (quasiprimary) fields. Typically one monomial produces a multiplet of spin $s$ conformal higher spin fields for all $s \geq 4$, they are distinguished by their anomalous dimensions (in $CFT_3$) or by their mass (in $AdS_4$). We sum over these multiplets and the spins to obtain "string type fields", one for each such monomial.
\end{abstract}
\vspace {2cm}
\it{Talk presented partially at SYMPHYS-XIV} \end{center} \begin{center}\it{Tzakhkadzor, Armenia, August 16-22, 2010}
\end{center}
\newpage

\section{From higher spin gauge fields to primitive string fields}
We define higher spin gauge fields 
\begin{equation}
h^{(s)}(z; a)
\label{1.1}
\end{equation}
in $D$ dimensional flat space $z\in \mathcal{R}_{D}, a \in \mathcal{T}_{D}(z)$ satisfying the constraints (valid also off shell)
\begin{eqnarray}
(a\partial_{a})h^{(s)}(z; a) = s \cdot h^{(s)}(z;a) \label{1.2}\\
\Box_{a}^{2} h^{(s)}(z;a) = 0 \label{1.3}
\end{eqnarray}
whose infinitesimal gauge transformations are
\begin{equation}
\delta_{\epsilon}h^{(s)}(z;a) = s \cdot (a\nabla) \epsilon^{(s-1)}(z;a)
\label{1.4}
\end{equation}
Derivatives on $z$ are denoted $\nabla$ and on $a$ by $\partial_{a}$. The trace of the gauge function $\epsilon^{(s-1)}$
vanishes 
\begin{equation}
\Box_{a}\epsilon^{(s-1)} = 0
\label{1.5}
\end{equation}

A free massless higher spin field satisfies Fronsdal's equation
\begin{eqnarray}
\mathcal{F}_{a}h^{(s)}(z;a) = 0 \label{1.6}\\
\mathcal{F}_{a} = \Box -(a \cdot\nabla)\ D_{a}, \Box = \nabla\cdot\nabla \label{1.7} \\
D_{a} = (\partial_{a}\cdot\nabla) -1/2 (a\cdot\nabla)\Box_{a}\quad \textnormal{(deDonder operator)} \label{1.8}
\end{eqnarray}
A very practical constraint is to impose deDonder gauge
\begin{equation}
D_{a}h^{(s)}(z;a) = 0
\label{1.9}
\end{equation}
 off shell. It corresponds to Feynman gauge in QED.

By Fourier transformation we obtain the operators 
\begin{eqnarray}
\hat{D_{a}} = (p\partial_{a}) -1/2 (ap)\Box_{a} \label{1.10}\\
\hat{\mathcal{F}}_{a} = p^2 -(ap)\hat{D}_{a} \label{1.11}
\end{eqnarray}
Therefore the propagator in momentum space and deDonder gauge is
\begin{equation}
G^{(s)}(p; a,b) = \frac{1}{p^2} A^{(s)}(a,b)
\label{1.12}
\end{equation}
where $A^{(s)}(a,b)$ is double traceless
\begin{equation}
\Box_{a}^{2}A^{(s)}(a,b) = \Box_{b}^{2} A^{(s)}(a,b) = 0
\label{1.13}
\end{equation}
This implies 
\begin{equation}
A^{(s)}(a,b) = A^{(s)}_1 \chi_{s}^{\lambda}(a,b) + A^{(s)}_2 a^{2}b^{2}\chi_{s-2}^{\lambda}(a,b)
\label{1.14}
\end{equation}
where $A_{1,2}^{(s)}$ are normalization constants and $\chi_{s}^{\lambda}(a,b)$ are expressed by Gegenbauer polynomials
\cite{GR} 
\begin{eqnarray}
\chi_{s}^{\lambda}(a,b) = (a^{2}b^{2})^{s/2}C_{s}^{\lambda}(\frac{(ab)}{(a^{2}b^{2})^{1/2}}) \label{1.15}\\
C_{-n}^{\lambda}(t) = 0 , n\in\mathcal{N}\label{1.16}\\
\lambda = \frac{1}{2} D -1 \label{1.17}
\end{eqnarray}

The generating function (synonymously 'primitive string field', PSF) is introduced by 
\begin{equation}
\Phi(z;a) = \sum_{s=0,even}^{\infty} h^{(s)}(z; a)
\label{1.18}
\end{equation}
In general we do not use any additional coefficient, a power $t^{s}$ is intrinsic in the s-th power of $a$, and can be mapped on a huge class of functions $f(s)$ by some integral or other kind of transform. If we define the normalization parameters $A_{1,2}
(a,b)$ in (1.14) to be
\begin{equation}  
A_1^{(s)} = \alpha_1 t^{s}, A_2^{(s)} = \alpha_2 t^{s} \label{1.19}
\end{equation}
then the propagator for the PSF is
\begin{eqnarray}
\Gamma_2 (p;a,b;t) =\qquad\qquad \qquad\qquad \nonumber\\
\frac {1}{2} (\alpha_1+a^2 b^2 t^2 \alpha_2)[(1-2(ab)t +a^2 b^2 t^2)^{-\lambda}+(1+2(ab)t +a^2 b^2 t^2)^{-\lambda}](p^2)^{-1}
\label{1.20}\\
= \int <\Phi(z_1;a)\Phi(z_2;b)> \exp\{-ip(z_1-z_2)\} d^{D}z_1 \label{1.21}
\end{eqnarray}

\section{The three-point functions and the corresponding currents}
\setcounter{equation}{0}

A general form (there are different forms!) of the cubic interaction for triplets of massless even spin fields was published 
\cite{MMR1}. This general form is 
\begin{eqnarray}
\mathcal{L}_3[h^{(s_1)},h^{(s_2)},h^{(s_3)}] = \sum_{Q_{12}Q_{23},Q_{31},\sum Q_{ij}=n}\nonumber\\
\frac{1}{Q_{12}!Q_{23}!Q_{31}!}\int dz_1dz_2dz_3\delta(z_1-z_2)\delta(z_1-z_3) \nonumber\\ Diff_{n}(Q_{12},Q_{23},Q_{31})
h^{(s_1)}(z_1;a_1)h^{(s_2)}(z_2;a_2)h^{(s_3)}(z_3;a_3) \label{2.1}
\end{eqnarray}
where $Diff_{n}$ is a differential operator built as follows:
\begin{enumerate}
\item $\nabla_1,\nabla_2,\nabla_3$ are coordinate derivatives, their total number is 
\begin{equation}
\Delta = s_1+s_2+s_3 -2n, n = \sum Q_{ij} \leq min\{s_1,s_2,s_3\}; \label{2.2}
\end{equation}
\item $\partial_{a_1}, \partial_{a_2}, \partial_{a_3}$ appear each of degree $s_{i}$ respectively;
\item $\Box_{a_1}, \Box_{a_2}, \Box_{a_3}$ are trace operations, at most one for each subscript;
\item $D_{a_1}, D_{a_2}, D_{a_3}$ are deDonder operators, at most one for each subscript.
\end{enumerate}

In deDonder gauge the terms with deDonder operators are absent, in this case there remain the 'leading terms' and the 'trace terms'.
If also the trace terms are set zero, we obtain espressions with only traceless transverse fields, whose degrees of freedom are too
small to describe off-shell fields. The leading terms of the differential operator are
\begin{eqnarray}
Diff_{n}(Q_{12},Q_{23}, Q_{31})_{lead} =\nonumber\\ (\nabla_1\partial_{a_3})^{s_3-n+Q_{12}}(\nabla_2\partial_{a_1})^{s_1-n+Q_{23}}(\nabla_3\partial_{a_2})^{s_2-n+Q_{31}}\nonumber\\
(\partial_{a_1}\partial_{a_2})^{Q_{12}}(\partial_{a_2}\partial_{a_3})^{Q_{23}}(\partial_{a_3}\partial_{a_1})^{Q_{31}}
\label{2.3}
\end{eqnarray}
The construction principle is obviously cyclic ordering. How we came to guess this principle is another story not to be told here.

It turns out that a cubic interaction $\mathcal{L}_3[\Phi_1,\Phi_2,\Phi_3]$ can be found from the cubic interactions of the HS fields
by an ansatz derived from string theory \cite{Sag},\cite{MMR2}. We find 
\begin{eqnarray}
\mathcal{L}_3[\Phi_1,\Phi_2,\Phi_3] = \int dz_1dz_2dz_3 \delta(z_1-z_2)\delta(z_1-z_3) \nonumber\\
\sum_{k,l} \rho_{k,l}\mathcal{A}_{k,l}\Phi_1(z_1;a_1)\Phi_2(z_2;a_2)\Phi_3(z_3;a_3)
\label{2.4}
\end{eqnarray}
where $k$ in $\{k,l\}$ denotes the number of different deDonder and $l$ the number of different trace operators. In a representaion 
using differences of coordinate derivatives and only leading and trace terms, there remain the three blocks with zero, two and three traces ($\nabla_{i}-\nabla_{j}= \nabla_{ij}$)
\begin{eqnarray}
\mathcal{A}_{k,l} = P(a_1,a_2,a_3)\exp W \times \mathcal{D}_{k,l} \label{2.5} \qquad\qquad\qquad\qquad\qquad\\
W = [(\partial_{a_1}\partial_{a_2}) +1](\partial_{a_3}\nabla_{12}) + [(\partial_{a_2}\partial_{a_3})+1](\partial_{a_1}\nabla_{23})
+ [(\partial_{a_3}\partial_{a_1}) +1](\partial_{a_2}\nabla_{31}) \label{2.6}\\
\mathcal{D}_{0,0} = 1\label{2.7}\qquad\qquad\qquad\qquad\qquad\qquad\\ \mathcal{D}_{0,2} = -[(\partial_{a_3}\nabla_{12})^2\Box_{a_1}\Box_{a_2} + (\partial_{a_1}\nabla_{23})^2 \Box_{a_2}\Box_{a_3} +(\partial_{a_2}\nabla_{31})^2\Box_{a_3}\Box_{a_1}]\label{2.8}\qquad\\
\mathcal{D}_{0,3} = (\partial_{a_1}\nabla_{23})(\partial_{a_2}\nabla_{31})(\partial_{a_3}\nabla_{12})\Box_{a_1}\Box_{a_2}\Box_{a_3}\label{2.9}\qquad\qquad\qquad\\
\rho_{0,0} =1, \rho_{0,2} = \frac{1}{16}, \rho_{0,3} = \frac{1}{32}\label{2.10}\qquad\qquad\qquad\qquad\qquad
\end{eqnarray}
where $P(a_1,a_2,a_3)$ projects on the subset $a_1=a_2=a_3 =0$.

From these expressions we can easily obtain expressions for the conserved currents by variation with respect to one field. For example by differentiation of the leading terms we get
\begin{eqnarray}
\mathcal{J}_{lead}[\Phi_1,\Phi_2](z;a) = P(a_1,a_2)\exp\{[(\partial_{a_1}\partial_{a_2}) +1](a\nabla_{12}) +[(\partial_{a_2}, a) +1](\partial_{a_1},(2\nabla_2+\nabla_1)\nonumber\\ - [(a,\partial_{a_1})+1] (\partial_{a_2},2\nabla_1+\nabla_2)\}\Phi_1(z_1;a_1)\Phi_2(z_2;a_2)\mid_{z_1=z_2 =z}\qquad\qquad
\label{2.11}
\end{eqnarray}

\section{Four-point functions}
\setcounter{equation}{0}
If we contract two currents of the same spin by a propagator we obtain a four-point function. According to whether the 
momentum $P$ carried by the propagator gives $P^2$ equal one of the Mandelstam variables $S, T, U$ we say that the four-point function is defined for the $S-, T-$, respectively $U$-channel. We will give the formulas here only for the $S$-channel, but
a complete four-point Green function contains the sum of all three channels.
Before we identify coordinates inside the two vertices we have six coordinate variables: $z_1$, $z_2$, $z_5$ in one vertex,
and $z_3$, $z_4$, and $z_6$ in the other vertex. The spin polarization vectors are denoted $a_{i}$ correspondingly. The propagator in deDonder gauge is in momentum space (see (1.20))
\begin{eqnarray}
\Gamma_2(P;a_5,a_6) = \frac{1}{2} (\alpha_1+(a_5)^2(a_6)^2 t^2\alpha_2)[\Omega_{+}(a_5,a_6) + \Omega_{-}(a_5,a_6)](P^2)^{-1} \label{3.1}\\
\Omega_{\pm}(a,b) = (1\pm2(ab)t + a^2b^2t^2)^{-\lambda}\label{3.2}
\end{eqnarray}
In coordinate space we obtain first for the higher spin fields themselves (we use the representation of the cubic interactions in terms of simple coordinate derivatives, denote $\partial_{a_{i}}$ by $\partial_{i}$, the parameters $n$ for the two vertices, respectively,
by $n_1$ and $n_2$, and use $s_5=s_6=s$) and taking into account only the leading terms at both vertices
\begin{eqnarray}
\sum_{Q_{i,j}}[Q_{12}!Q_{25}!Q_{51}!Q_{34}!Q_{46}!Q_{63}!]^{-1}\qquad\qquad\nonumber\\
(\partial_5\nabla_1)^{s-n_1+Q_{12}}(\partial_1\nabla_2)^{s_1-n_1+Q_{25}}(\partial_2\nabla_5)^{s_2-n_1+Q_{51}}\nonumber\\
(\partial_6\nabla_3)^{s-n_2+Q_{34}}(\partial_3\nabla_4)^{s_3-n_2+Q_{46}}(\partial_4\nabla_6)^{s_4-n_2+Q_{63}}\nonumber\\
(\partial_1\partial_2)^{Q_{12}}(\partial_2\partial_5)^{Q_{25}}(\partial_5\partial_1)^{Q_{51}}\quad\quad\quad\nonumber\\
(\partial_3\partial_4)^{Q_{34}}(\partial_4\partial_6)^{Q_{46}}(\partial_6\partial_3)^{Q_{63}}\quad\quad\quad\nonumber\\
A^{(s)}(a_5,a_6)((z_5-z_6)^2)^{-\lambda}\prod_{i=1}^{4}h^{(s_{i})}(z_{i}; a_{i}) \quad\quad\label{3.3}\\
Q_{12}+Q_{25}+Q_{51} = n_1, Q_{34}+Q_{46} +Q_{63} = n_2 \label{3.4}
\end{eqnarray}
with $A^{(s)}(a_5,a_6)$ from (1.12)-(1.14).

Under transformation to momenta $\nabla_{i},i\in {1,2,3,4}$ goes into $p_{i}$ (constant factors are neglected) and
momentum conservation gives
\begin{equation}
p_1+p_2 = P = p_3 + p_4 \label{3.5}
\end{equation}
Further $[z_{56}^2]^{-\lambda}$ goes into $(P^2)^{-1}$. Moreover we can replace $\nabla_5$ and $\nabla_6$
by $P$. The homogeneity of the four-point function in all momenta is $n_1+n_2-2$.
There is a tensor in the momentum vector $P$ of the rank 
\begin{equation}
\Pi_{\mu_1\mu_2...\mu_{N}}^{(N)} = \bigotimes_{i=1}^{N} P_{\mu_{i}}, \label{3.6}
\end{equation}
contracted to
\begin{equation}
(\partial_2 P)^{s_2-n_1+Q_{51}}(\partial_4 P)^{s_4-n_2+Q_{63}}
\label{3.7}
\end{equation}
implying 
\begin{equation}
N = N_1 + N_2,\quad N_1 = s_2-n_1+Q_{51}, \quad N_2 = s_4-n_2+Q_{63} \label{3.8}
\end{equation}
When we decompose this tensor (3.6) into a traceless part (using a Gegenbauer polynomial, say, making it unique this way) and a trace part, the latter one is of a polynomial dependence in $P$ since $(P^2)^{-1}$ cancels, and therefore local in coordinate space. Since the whole four-point function is linearly gauge invariant, the gauge variations of local and nonlocal parts must both be local.

Now we turn to the four-point functions of the PSF's. We want to evaluate only the leading terms (no traces, but traces must be kept free
in deDonder gauge!). We start from (2.5)-(2.7) and (2.11) so that we get
\begin{equation}
\exp[W_{125} + W_{346}] \prod_{i=1}^{4}\Phi_{i}(z_{i};a_{i}) <\Phi_5(z_5;a_5)\Phi_6(z_6;a_6)> \label{3.9}
\end{equation}
where
\begin{equation}
W_{125} = [(\partial_1\partial_2)+1](\partial_5\nabla_1)+ [(\partial_2\partial_5)+1](\partial_1\nabla_2)
+[(\partial_5\partial_1)+1](\partial_2\nabla_5) \label{3.10}
\end{equation}
and $W_{346}$ analogously. To a certain degree the structure of (3.9) allows us to do some reduction (evaluation).
Namely it is possible to apply the operators $\partial_5, \partial_6$ to the spin polarization factor in the propagator, and independently the gradients $\nabla_5, \nabla_6$ to the coordinate factor. This can be done independently since in deDonder gauge the propagator factorizes into these two parts.

First we extract $\nabla_5, \nabla_6$ (being equivalent to $-\nabla_5$ because it acts only on  $(z_5-z_6))$ from the exponent operators $W_{125}, W_{346}$ (see (3.10))
\begin{eqnarray}
\exp\langle(\partial_2\nabla_5)[(\partial_5\partial_1)+1] - (\partial_4\nabla_5)[(\partial_6\partial_3) +1] \rangle= \nonumber\\
\sum_{N_1,N_2} \frac{1}{N_1!N_2!}(\frac{\partial}{\partial \rho_1})^{N_1} (\frac{\partial}{\partial \rho_2})^{N_2}
\exp\langle [(\partial_5\partial_1)+1] \rho_1 +[(\partial_6\partial_3)+1] \rho_2\rangle \mid_{\rho_1 =\rho_2 = 0}\nonumber\\
\times (-1)^{N_2}(\partial_2\nabla_5)^{N_1}(\partial_4\nabla_5)^{N_2}\qquad\qquad\qquad
\label{3.11}
\end{eqnarray}
We can then perform the differentiations $\nabla_5$. The differentiations $\partial_5,\partial_6$ are all contained in the exponent. The remaining exponent is expanded as follows
\begin{eqnarray}
\exp\{(X\partial_5) + (Y\partial_6) +U +V +\rho_1 +\rho_2\}\label{3.12}\qquad\\
X(\rho_1) = [(\partial_1\partial_2) +1] \nabla_1 +(\partial_1\nabla_2)\partial_2 + \rho_1\partial_1 \label{3.13}\\
Y(\rho_2) = [(\partial_3\partial_4) +1] \nabla_3 + (\partial_3\nabla_4)\partial_4 +\rho_2\partial_3 \label{3.14} \\
U = (\partial_1\nabla_2), V = (\partial_3\nabla_4) \label{3.15}\qquad
\end{eqnarray}
and we perform the differentiations on $a_5,a_6$ using $\Omega_{\pm}$ from (3.2) to obtain
\begin{eqnarray}
\exp[W_{125} + W_{346}]<\Phi_5\Phi_6> \prod_{i = 0}^{4}\Phi_{i}(z_{i};a_{i}) = \qquad\qquad\nonumber\\
\sum_{N_1,N_2}\frac{1}{N_1 ! N_2 !} (\frac{\partial}{\partial\rho_1})^{N_1}(\frac{\partial}{\partial\rho_2})^{N_2}\exp(\rho_1+\rho_2)\qquad\qquad\nonumber\\
\frac{1}{2}(\alpha_1 + \alpha_2 X(\rho_1)^2 Y(\rho_2)^2 t^2)\{\sum_{\pm}\Omega_{\pm}(X(\rho_1),Y(\rho_2))^{-\lambda}\}\qquad
\nonumber\\(-1)^{N_2}(\partial_2 \nabla_5)^{N_1} (\partial_4\nabla_5)^{N_2}[(z_5-z_6)^2]^{-\lambda}\exp[U+V]\prod_{i=0}^{4} \Phi_{i}(z_{i};a_{i})\mid_{a_{i} = 0}
\label {3.16}
\end{eqnarray}

\section{AdS/CFT correspondence for PSF's}
\setcounter{equation}{0}
The AdS viewpoint of these PSF's can be comfortably discussed for the "standard models" of AdS/CFT correspondence, whose CFT 
is the $O(N)$ invariant sector of the O(N) vector sigma model in the free case or the critical interacting case at the conformal fixed point \cite {KP}. It possesses a scalar-isovector field $\phi_{i}(x)$ from which we can construct two classes of conserved (free case) or almost conserved (critical case) currents which in the free case are constructed from bilinear bilocal fields by operator product expansion
\begin{eqnarray}
(1)\quad b(x, y) = N^{-1/2}\sum_{i}\phi_{i}(x)\phi_{i}(y) \label{4.1}\\
(2)\quad b_{A}(x,y) = N^{-1/2}\sum_{i,j}\gamma_{A}^{i,j}\phi_{i}(x)\phi_{j}(y) \label{4.2}
\end{eqnarray}
where $\gamma_{A}^{i,j}$ is an (antisymmetric) matrix of the adjoint representation of $so(N)$ whose commutators have  totally antisymmetric structure constants $f_{ABC}$. There is a natural embedding of $O(N)$ into $O(N+1)$ entailing the embedding of the Lie algebras of both groups whose inverse is a projection of linear spaces. We therefore need not specify these Lie algebras for a particular $N$ which entails the possibility for a $1/N$ expansion.         

The number $N$ can in principle vary over all natural numbers. For $N=1$ the symmetry group is $O(N) = Z_2$, and this sigma model is the Ising model. For large $N$ these models can be evaluated by $1/N$ expansion. All these models have a free and a critical version, the latter versions define particular conformal field theories. These critical versions can be used to define universality classes, which implies that sets of models, e. g. the one-component Ginzburg-Landau model, has the same critical behaviour and the same conformal field theory as the Ising model, or for the two-component Ginzburg-Landau model the same conformal field theory results at the critical point as for the $O(2)$ sigma model. Each universality class is characterized by its space dimension, the number of real components and the symmetry group. The critical (conformal) theories are characterized by their operator product expansions, for their evaluation we have so far only the $1/N$ expansions for large $N$. However, we will see that a numerical study of the anomalous dimension of the fundamental $N$-vector of fields  $\phi_{i}$ (which are also called the "order parameters") allows us a first order formula of the HS field mass squared for any $N$. 

By operator product expansion it is possible to define currents for the free conformal field theory
\begin{eqnarray}
J^{(s)}(y;a) = N^{-1/2}\sum_{r=0}^{s/2}\sum_{n=0}^{s-2r}A_{r,n}^{(s)}(a^2)^{r}(a\nabla)^{n}[\nabla_{\nu_1}...\nabla_{\nu_{r}}]
\phi_{i}(y)\nonumber\\(a\nabla)^{s-n-2r}[\nabla_{\nu_1}...\nabla_{\nu_{r}}] \phi_{i}(y)\qquad ( s \quad\textnormal{is even})\label{4.3}\\
J_{A}^{(s)} (y;a) = N^{-1/2}\sum_{r=0}^{(s-1)/2}\sum_{n=0}^{s-2r}A_{r,n}^{(s)}(a^2)^{r}(a\nabla)^{n}
\gamma_{A}^{i,j}[\nabla_{\nu_1}...\nabla_{\nu_{r}}]\phi_{i}(y)\nonumber\\(a\nabla)^{s-n-2r}[\nabla_{\nu_1}...\nabla_{\nu_{r}}] \phi_{j}(y)\qquad(s \quad \textnormal{is odd}) \label{4.4}\\
A_{r,n}^{(s)} = \frac{(-1)^{s}}{2^{r}r!n!(s-n-2r)!} \frac{(\delta)_{s-r}}{(\delta)_{r+n} (\delta)_{s-n-r}}\qquad\qquad \label{4.5}\\
\delta = \frac{1}{2}d -1 = \frac{1}{2} \quad \textnormal{for d =3} \qquad\qquad\qquad\label{4.6}
\end{eqnarray}
These currents of tensorial rank $s$ are conserved, traceless, and have conformal dimension $d+s-2 = s+1$. We have made use of the free field equations for $\phi$ that imply also
\begin{equation} 
\frac{1}{2} \Box [ \phi(y)\phi(y)] = \nabla_{\nu}\phi(y) \nabla_{\nu}\phi(y) \label{4.7}
\end{equation}
In turn an operator product expansion yields the inverse to (4.3) with expansion around the center $y$ of the interval $(x_1,x_2)$
\begin{equation}
a = x_1-y = y-x_2 \label{4.8}
\end{equation}
giving
\begin{equation}
b(x_1,x_2) = \sum_{M=0}^{\infty}\sum_{s=0,\ even}^{M}\sum_{t=0}^{\infty}B_{M,s,t}(a\partial_{y})^{M-s}(\frac{1}{2}a^2\Box)^{t}
J^{(s)}(y;a) \label{4.9}
\end{equation}
In this operator product expansion (which has an analogous counterpart for the currents of the O(N) Lie algebra) 
no non-conserved "twisted currents" $J^{(s,t)}(y;a)$ of dimension
\begin{equation}
\delta_{s,t} = d + s +2t -2 = s +2t +1 , t> 0 \label{4.10}
\end{equation}
arise. Finally we give the two-point function for such conserved currents
\begin{eqnarray}
\langle J^{(s)}(y_1;a) J^{(s)}(y_2;b)\rangle = \mathcal{N}_{s} (y_{12}^2)^{-s-1}\frac{s!}{2^{s}(\delta)_{s}}(\mid a\mid\mid b\mid)^{s}C_{s}^{\delta} (\xi) \label{4.11} \\
\mathcal{N}_{s} = \frac{2 s!(2\delta+s-1)_{s}}{(\delta)_{s}^2}\qquad\qquad\qquad \label{4.12} \\
\xi = \frac{1}{\mid a\mid\mid b \mid}[2\frac{(ay_{12})(by_{12})}{y_{12}^2}- (ab)]\qquad\qquad \label{4.13}
\end{eqnarray}
where $C_{s}^{\delta}(\xi)$ is the Gegenbauer polynomial.

Before we turn to the AdS case, we consider the critical $O(N)$ models. The main change is that the scalar field $J^{(0)}$ is now replaced by the conformally dual $\alpha$ field, and that all dimensions obtain an anomalous part except the spin-two current
(energy-momentum tensor) and the spin-one Lie algebra current (O(N) charge currents). Only these two types of currents remain conserved. For large $N$ all anomalous dimensions can be $1/N$ expanded
\begin{equation}
\eta = \sum_{r=1}^{\infty}\frac{\eta_{r}}{N^{r}} \label{4.14}
\end{equation}
and the same holds for the critical coupling constant 
\begin{equation}
z = \sum_{r=1}^{\infty}\frac{z_{r}}{N^{r}} \label{4.15}
\end{equation}
The majority of anomalous dimensions for this model have first been calculated by the group of authors \cite{VPK}. The anomalous dimensions for the vector currents $J^{(s)}$ for the critical case have been calculated in \cite{LR}. From these expressions 
the masses squared of the higher spin fields (with gauge symmetry broken) have been derived in \cite{R}.

The coupling parameter of the term $\alpha \phi^2$ is $z^{1/2}$. Since $\alpha$ and $\phi^2$ are conformally dual and $\phi^2$
couples trivially to itself, $z_1$ is contained in the intertwiner of the dual pair and is thus purely kinematical. More
interesting is that the operator product expansion approach using skeleton graphs only, determines the part $\kappa$ in the anomalous dimension $\delta(\alpha)$ of $\alpha$ 
\begin{equation}
\delta(\alpha)= 2- 2\eta(\phi) -2\kappa \label {4.16}
\end{equation} 
by a consistency argument. Going through the details of the calculation of the anomalous dimensions \cite{LR} we see that among the critical parameters
only $\eta(\phi)$ 
\begin{equation}
\delta(\phi) = \frac{d}{2} - 1 +\eta(\phi) = \frac{1}{2} + \eta(\phi), \quad d = 3 \label{4.17}
\end{equation}
comes from outside. In a first order calculation in $N^{-1}$ we can therefore replace consistently $\eta_1(\phi)/N$
by a numerically determined parameter $\eta_{num}$ for the critical anomalous dimension of the $O(N)$ vector field $\phi$  
(the "order parameter" of the $O(N)$ models), taken from statistical mechanics calculations or even experiment. This parameter is small for all $N$. Then in the $1/N$ calculations to first order replace
the left hand expressions by the right hand ones (all at $d=3$) in
\begin{eqnarray}
\kappa_1 \rightarrow -3 \eta_1(\phi) \label{4.18}\\
z_1 \rightarrow  \frac{3}{4\pi^2}\eta_1(\phi) \label{4.19}
\end{eqnarray}
Collect all factors of $\eta_1(\phi)$, sum them and then replace 
\begin{equation}
N^{-1}\eta_1(\phi)\rightarrow \eta_{num}(\phi) \label{4.20}
\end{equation}
Inserting these into the anomalous dimensions of the $O(N)$ scalar currents of even spin $s \geq 2$ as calculated in \cite{LR}, we get
\begin{equation}
\eta(J^{(s)}) = \frac{4(s-2)}{2s-1}\eta_{num}(\phi) + higher order terms\label{4.21}
\end{equation}
which determines the linear term in $\eta_{num}(\phi)$ of the mass squared of $h^{(s)}$
\begin{equation}
m(s)^2 = 4(s-2)\eta_{num}(\phi)+ higher order terms \label{4.22}
\end{equation}
Values for $\eta_{num}(\phi)$ at small $N$ can be found in the literature (what we call $\eta$ corresponds to $\eta/2$ for most authors in statistical mechanics, who deal with correlation functions and not fields). They are in general small indeed, $N=1: 0,012$ (Ising), $N=3: 0,035$ (Heisenberg), $N=\infty: 0$ (spherical)). These formulae (4.21), (4.22) are supposed to be valid for all $N$ and all $s$, but how good they are can only be tested numerically.

The bilinear bilocal scalar field (4.1)
can be decomposed in this interacting case into currents and the scalar field $\alpha(x)$ by operator product expansion as follows              
\begin{eqnarray}
b(x_1,x_2) = g_{\alpha}\int dx K_{\alpha}(x_1,x_2; x)\alpha(x) \qquad\qquad\qquad \nonumber\\
+ \sum_{s\geq 2, even}^{\infty}g_{s} \int dx K^{(s)}(x_1,x_2,x;\partial_{a})J^{(s)}(x;a) \qquad\qquad \nonumber\\
+\sum_{s\geq 0, even}^{\infty}\sum_{t=1}^{\infty}g_{s,t}\int dx K^{(s,t)}(x_1,x_2,x;\partial_{a})J^{(s,t)}(x;a) \label{4.23}
\end{eqnarray}
The integral kernels $K$ can be represented as convolution kernels depending on the dimension $\delta$ of $\phi$ which now contains an anomalous part. The coupling constants $g_{s}$ are known including their first order anomalous part.

Now we turn to the subject of this article: the PSF's. Of course a PSF on Minkowski space can be lifted \cite{R2} in a straightforward way to a PSF in AdS space such that
\begin{equation}
\sum_{s\geq 2,\ even}^{\infty}g_{s}J^{(s)}(\vec x;\vec a)\quad \textnormal{goes into} \sum_{s\geq 2,\ even}^{\infty}g_{s}h^{(s)}(z;a)
\label{4.24}
\end{equation}
and similarly in all the other cases of Minkowski conformal currents. If the currents are conserved (implying they have canonical dimension)
the higher spin fields have mass zero and can be gauge transformed. In the case of the critical O(N) sigma model, from all the
currents mentioned only the spin-1 Lie algebra current and the O(N) scalar spin-2 current are conserved and lead to gauge degrees
of freedom for the AdS fields. In these O(N) examples the representations involved in the PSF are pairwise inequivalent. Therefore one can easily define a map from CFT to AdS field theory adding up the Dobrev kernels \cite{Dob} correspondingly, understanding that each of these kernels annihilates objects of an inequivalent representation. A practical way to do this is used later.

Since all propagators for higher spin fields on AdS are known \cite{LMR}, the propagators for PSF's can in principle be written down. However, for the PSF's one better goes another way. We use Dobrev's kernel and apply it to the conformal conserved currents $J^{(s)}$ of the free O(N) sigma model. They are made traceless and have dimension 
\begin{equation}
\Delta_{s} = d+s-2 \label{4.25}
\end{equation}
so that PFS's on AdS space have the structure of power series.
From \cite{R2} we obtain for Dobrev's kernel \cite{Dob}
\begin{equation}
K_{\Delta_{s}}^{(s)}(z,\vec x; b, \vec a) = \frac{z_0^{d-2}}{((z-\vec x)^2)^{d+s-2}}
\langle\{ \sum_{\beta=0}^{d}\sum_{i=1}^{d}\{b_{\beta}r(z-\vec x)_{\beta,i}\vec a_{i}\}^{s}- \textnormal{traces}\rangle
\label{4.26} 
\end{equation}
where
\begin{equation}
r(z-\vec x)_{\beta,i} = 2\frac{(z_{\beta}-\vec x_{\beta})(z_{i}-\vec x_{i})}{(z-\vec x)^2} -\delta_{\beta,i},\qquad \vec x_0 = 0
\label{4.27}
\end{equation}
This convolution operator can be made to include a projection on symmetric tensors of rank $s$ if we replace
\begin{equation}
\vec a_{i} \rightarrow \vec \partial_{a_{i}} \label{4.28}
\end{equation}
and set all remaining $a$ after the differentiation equal zero. Then $K^{(s)}$ acts on $J^{(s)}$ only. Moreover since $a_0=0$
is always understood in the currents, a differentiation w.r.t. this variable is void and can always be included. Thus we 
may use the partial derivative $\partial_{a_{\alpha}}$ with $\alpha$ running over $d+1$ labels. 

After an appropriate normalization we obtain, using Gegenbauer polynomials $C_{s}^{\lambda}$, the traceless HS field
\begin{eqnarray}
h^{(s)}(z;b) = \int d\vec x K_{\Delta_{s}}^{(s)}(z,\vec x;b, \partial_{a})J^{(s)}(\vec x; \vec a) \nonumber\\
= \int d\vec x(\frac{z_0}{(z-\vec x)^2})^{d-2} \zeta^{s}C_{s}^{\lambda }(t) J^{(s)}(\vec x;\vec a)
\label{4.29}
\end{eqnarray}
where
\begin{eqnarray}
t = \frac{1}{(b^2 \Box_{a})^{1/2}}\langle 2\frac{(b,z-\vec x)(z-\vec x,\partial_{a})}{(z-\vec x)^2} - (b\partial_{a})\rangle \label{4.30}\\
\zeta = \frac{(b^2 \Box_{a})^{1/2}}{(z-\vec x)^2}\qquad\qquad\qquad\label{4.31}
\end{eqnarray}
The expression
\begin{equation}
\zeta^{s}C_{s}^{\lambda} \label{4.32}
\end{equation}
is a polynomial of (integral) degree $\frac{s}{2}$ of the variable $b^2\Box_a$.

Now we introduce the traceless PSFs on AdS by summing
\begin{equation}
\Psi(z; b) = \sum_{s\geq 0,\ even}^{\infty}h^{(s)}(z;b) \label{4.33}
\end{equation}
On the r.h.s. of (4.29) we can introduce a double sum because of the termwise orthogonality
\begin{eqnarray}
\Phi(\vec x; \vec a) = \sum_{s\geq 0,\ even}^{\infty} J^{(s)}(\vec x; \vec a)\label{4.34}\qquad\qquad\qquad \\
\Psi(z; b) = \frac{1}{2}\int d\vec x \langle \frac{z_0}{(z-\vec x)^2}\rangle^{d-2} [\Omega_{-}(z,\vec x; b, \partial_{a})
+\Omega_{+}(z,\vec x;b, \partial_{a})] \Phi(\vec x; \vec a) \label{4.35}
\end{eqnarray}
The functions $\Omega$ arise through generating functions of Gegenbauer polynomials
\begin{equation}
\Omega_{\pm}(z, \vec x; b, \partial_{a}) = (1 \pm 2t\zeta + \zeta^2)^{-\lambda} \label{4.36}
\end{equation}
Note that the dimension $\delta$ of the free $O(N)$ vector field in $d$ dimensional Minkowski space ($d/2 -1$) is not the same as $\lambda$ in the Gegenbauer polynomials ($\lambda = (d+1)/2-1$) in (4.29) - (4.36).

In the case of the critical O(N) model traceless PSFs can of course also be defined but explicit summing over the spin $s$ is not 
elementary any more. The subsequent section is devoted to these.

\section{String type fields for the critical $O(N)$ models}
\setcounter{equation}{0}
In Section 1 we introduced primitive string fields by summing over the HS fields belonging to the lowest Regge trajectory.
PSF and Regge trajectory can be considered as two different mathematical descriptions of the same object. This changes now.
We consider the AdS/CFT image of a "derivation module": we start from one O(N) scalar product of two vector fields $\phi$ and
apply arbitrary many derivations $(a\partial)$ to it. Linear combinations of these higher derivations span a module. In this module we must select quasiprimary (conformal) fields (which was trivial in the PSF case). For the identification of the quasiprimary fields we need the anomalous dimensions. These have to be calculated by $1/N$ expansion. Besides the currents of the critical O(N) obtained from the single scalar product of vectors, we could also use higher powers of such scalar products of two vectors, powers of the scalar field $\alpha$, and mixed products of these as starting basis to which the derivations are applied. In the case of the products of $\alpha$ itself, we can again rely on work done with Klaus Lang in 1993-94 \cite{LR1}, \cite{LR2}, and therefore we can present here some results. All string type fields obtained for the O(N) scalar sector of the critical O(N) vector model (the "standard" CFT) are constructed from totally symmetric tensors, depending therefore only on one vector $a$ of the tangential space of $AdS_{4}$.

We consider the scalar field $\alpha(x)$ of conformal dimension
\begin{equation}
\delta_{\alpha} = 2 +\sum_{k=1}^{\infty}\frac{\eta_{k}(\alpha)}{N^{k}} \label{5.1}
\end{equation}
Defining normal products for scalar conformal fields $A(x)$ and $B(x)$ by 
\begin{equation}
A(x)B(y) = ((x-y)^2)^{1/2(\delta_{C}-\delta_{A}-\delta_{B})}[:A(x)B(x): + O(x-y)],\quad C(x) = :A(x)B(x):
\label{5.2}
\end{equation}
we start from the basis $:\alpha(x)^{r}:, r\in\{2, 3, 4,...\}$
and apply $s$ derivations
\begin{eqnarray}
\psi_{\{l_1,l_2,..l_{r}\}}(x;a) = :(a\partial)^{l_1}\alpha(x)(a\partial)^{l_2}\alpha(x)...(a\partial)^{l_{r}}\alpha(x): - \textnormal{trace terms}\nonumber\\
l_1+l_2+...l_{r} = s \qquad\qquad\qquad
\label{5.3}
\end{eqnarray}
where the powers $l_{i}$ are ordered conveniently
\begin{equation}
l_1 \geq l_2\geq l_3 ...\geq l_{r} \geq 0
\label{5.4}
\end{equation}
Then we consider linear combinations
\begin{equation}
\sum_{\textnormal{partitions of s of length r}}C_{\{l_1,l_2,...l_{r}\}} \psi_{\{l_1,l_2,...l_{r}\}}(x)
\label{5.5}
\end{equation}
forming the module $\mathcal{M}(r,s)$. Not all of its vectors are quasiprimary (conformal), however. Namely, if it is possible to factor one derivative
$(a\partial)$ we have a "derivative field" (better would be the term "integrable field"). The dimension of $\mathcal{M}(r,s)$ is equal the number of ordered partitions of s of length r, which we denote $p_{r}(s)$. The dimension of the subspace $\mathcal{M}(r,s)_{conf}$ spanned by quasiprimary fields (we call it the "multiplicity") is then by simple recursion
\begin{equation}
dim\{\mathcal{M}(r,s)_{conf}\} = p_{r}(s) -p_{r}(s-1).
\label{5.6}
\end{equation}
The conformal dimension of the quasiprimary fields in this subspace is (by (5.1))
\begin{equation}
\delta_{r,s,k} = 2r +s +O(\frac{1}{N})
\label{5.7}
\end{equation}
where k counts basis elements (conformal fields) of the multiplet $\mathcal{M}(r,s)_{conf}$ that by definition must have well defined anomalous dimensions.
These are eigenvalues of some matrix that can be determined explicitly, and are evaluated by a spectral decomposition algorithm. Conformal fields with different conformal dimensions produce orthogonal states in Hilbert space. 

It turns out that at the order $1/N$ some of the anomalous dimensions may 
coincide. In \cite{LR1} we have studied the case $(r,s) = (3,6)$, where such a situation arises, in detail. We could specify two different quasiprimary fields which at space dimension $d=3$ have the same conformal dimension to the order $1/N$, but by analytic continuation in $d: 2\leq d\leq 4$, each has a conformal dimension analytic in $d$, with both functions intersecting at $d=3$. All cases familiar from atomic physics may occur at the next order $1/N^2$, e.g. level crossing etc. The technique of determining the anomalous dimensions is based in \cite{LR1} on a cohomological analysis 
of the direct sum $\sum_{\oplus r,s}\mathcal{M}(r,s)$. For illustration we give a table of multiplicities below (from \cite{LR1}).
\begin{eqnarray}
\begin{array}{ccccccccccccc}   
r= &1&2&3&4&5&6&7&8&9&10&11&12 \\
s= & & & & & & & & & & & & \\
0:  &1&1&1&1&1&1&1&1&1&1&1&1 \\
1:  &0&0&0&0&0&0&0&0&0&0&0&0 \\
2:  &0&1&1&1&1&1&1&1&1&1&1&1 \\
3:  &0&0&1&1&1&1&1&1&1&1&1&1 \\
4:  &0&1&1&2&2&2&2&2&2&2&2&2 \\
5:  &0&0&1&1&2&2&2&2&2&2&2&2 \\
6:  &0&1&2&3&3&4&4&4&4&4&4&4 \\
7:  &0&0&1&2&3&4&4&4&4&4&4&4 \\
8:  &0&1&2&4&5&6&6&7&7&7&7&7 \\
9:  &0&0&2&3&5&6&7&7&8&8&8&8 \\
10: &0&1&2&5&7&9&10&11&11&12&12&12 \\
11: &0&0&2&4&7&9&11&12&13&13&14&14 \\
12: &0&1&3&7&10&14&16&18&19&20&20&21 \\
\end{array}
\end{eqnarray}
\quad Table of multiplicities (5.6) for $r\leq 12, s \leq 12$: To the right of the diagonal $r=s$ the multiplicities are constant in $r$,
for $r=2$ only even $s$ occur as in the case of the standard PSF.  

By AdS/CFT correspondence we obtain a symmetric traceless tensor field  $h_{k}^{(r,s)}(z;a)$ of rank (spin) $s$ with the conformal dimension (5.7) and a corresponding mass squared for each $r\in\{2, 3, 4,...\}$ and $k$, where the degeneracy label $k$ runs over $p_{r}(s) - p_{r}(s-1)$ values. These fields are also considered to produce by simple application to a Hilbert space ground state pairwise orthogonal states. First we sum over these degeneracy labels with coefficients $\gamma_{k}^{(r,s)}$
\begin{equation}
\hat h^{(r,s)}(z;a) = \sum_{k=1}^{p_{r}(s)-p_{r}(s-1)}\gamma_{k}^{(r,s)}h^{(r,s)}_{k}(z;a)
\label{5.9}
\end{equation}
that may be normalized in some fashion. Then we introduce the string type field 
\begin{equation}
\Phi_{r}(z;a) = \sum_{s=0}^{\infty}\hat h^{(r,s)}(z;a)
\label{5.10}
\end{equation}
Here the orthogonality of all higher spin fields of fixed $r$ is crucial.

Needless to say that applications of these string type fields to string theory are an unsolved issue.

\section{Conclusion}
\setcounter{equation}{0}
Apart from the open problem of applying these string type fields, some quantum field theoretical problems
remain unsolved. The scalar product of two vector fields is dual to the scalar field $\alpha$. Though derivation
modules produced from powers $:(\vec \phi(x)\vec \phi(x))^{r}:$ produce conformal fields with different conformal 
dimensions
\begin{equation}
\delta_{(\phi^2)^{r}} = r + s + O(\frac{1}{N})
\label{6.1}
\end{equation}
the duality must have consequences relating (6.1) with (5.7). These are not yet investigated.

On the other hand mixed symmetry higher spin fields pose serious problems. First the propagator of such fields 
is unknown to our knowledge. Second \cite{Brink} in general a whole multiplet of conformal fields in $CFT_{d}$ is needed for the lift to one gauge covariant higher spin field of $AdS_{d+1}$. Most interesting would be a standard model to deal with such cases, allowing for a free field version and a perturbatively accessible critical version.  

\section{Acknowledgements}
This work and the meeting in Tsakhkadzor were made possible by Alexander von Humboldt-Foundation grants. We profited a lot from the lasting collaboration with the colleagues M. Manvelyan and K. Mkrtchyan from Yerevan Physics Institute.

\end{document}